\documentclass[pdflatex,iicol,sn-mathphys-num]{sn-jnl}

\usepackage{graphicx}%
\usepackage{multirow}%
\usepackage{amsmath,amssymb,amsfonts}%
\usepackage{amsthm}%
\usepackage{mathrsfs}%
\usepackage[title]{appendix}%
\usepackage{xcolor}%
\usepackage{textcomp}%
\usepackage{manyfoot}%
\usepackage{booktabs}%
\usepackage{algorithm}%
\usepackage{algorithmicx}%
\usepackage{algpseudocode}%
\usepackage{listings}%
\usepackage[normalem]{ulem}%

\usepackage{enumerate}

\theoremstyle{thmstyleone}%
\theoremstyle{thmstyletwo}%

\theoremstyle{thmstylethree}%

\begin{document}

\title[Article Title]{Crust (Unified) Tool for Equation-of-state Reconstruction (CUTER) v2}


\author[1]{\fnm{Philip J.} \sur{Davis}}\email{davis@lpccaen.in2p3.fr}
\equalcont{These authors contributed equally to this work.}

\author[2]{\fnm{Hoa} \sur{Dinh Thi}}\email{hoa.dinh@rice.edu}
\equalcont{These authors contributed equally to this work.}

\author*[3]{\fnm{Anthea F.} \sur{Fantina}}\email{anthea.fantina@ganil.fr}
\equalcont{These authors contributed equally to this work.}

\author[1]{\fnm{Francesca} \sur{Gulminelli}}\email{gulminelli@lpccaen.in2p3.fr}
\equalcont{These authors contributed equally to this work.}

\author[4,5]{\fnm{Micaela} \sur{Oertel}}\email{micaela.oertel@astro.unistra.fr}
\equalcont{These authors contributed equally to this work.}

\author[6]{\fnm{Lami} \sur{Suleiman}}\email{lsuleiman@fullerton.edu}
\equalcont{These authors contributed equally to this work.}

\affil[1]{\orgdiv{Université de Caen Normandie}, \orgname{ENSICAEN, CNRS/IN2P3, LPC Caen UMR6534}, \orgaddress{\city{Caen}, \postcode{14000}, \country{France}}}

\affil[2]{\orgdiv{Department of Physics and Astronomy}, \orgname{Rice University}, \orgaddress{\street{6100 Main Street}, \city{Houston}, \postcode{77251-1892}, \state{Texas}, \country{USA}}}

\affil*[3]{\orgdiv{Grand Acc\'el\'erateur National d'Ions Lourds (GANIL)}, \orgname{CEA/DRF - CNRS/IN2P3}, \orgaddress{\street{Boulevard Henri Becquerel}, \city{Caen}, \postcode{14076}, \country{France}}}

\affil[4]{\orgdiv{Observatoire astronomique de Strasbourg}, \orgname{CNRS, Université de Strasbourg}, \orgaddress{\street{11 rue de l'université}, \city{Strasbourg}, \postcode{67000}, \country{France}}}

\affil[5]{\orgdiv{LUX}, \orgname{CNRS, Observatoire de Paris}, \orgaddress{\street{5 place Jules Janssen}, \city{Meudon}, \postcode{92915}, \country{France}}}

\affil[6]{\orgdiv{Nicholas and Lee Begovich Center for Gravitational Wave Physics and Astronomy}, \orgname{California State University Fullerton}, \orgaddress{\city{Fullerton}, \postcode{92831}, \state{California}, \country{USA}}}


\abstract{The equation of state (EoS) is a needed input to determine the neutron-star global properties and to relate them. It is thus important to provide consistent and unified EoSs to avoid possible biases in the analyses coming from the use of inconsistent EoSs. 
We propose a numerical tool, \texttt{CUTER}, allowing the user to consistently match a nuclear-physics informed crust to an arbitrary higher density EoS. We present here the second version of this tool, \texttt{CUTER v2}.
Two functionalities are available with the \texttt{CUTER v2} tool, allowing the user to reconstruct either the whole (outer and inner) crust, or the outer crust only.
We show that the code, that has been tested and validated for use by the astrophysical
community, is able to efficiently perform both tasks, allowing the computation of neutron-star global properties in a consistent way.
}


%
%
%

\keywords{dense matter, equation of state, gravitational waves, neutron stars}

\maketitle

\section{Introduction}
\label{sec:intro}

The first multi-messenger detection of a binary neutron star (NS) merger, the event GW170817 \cite{Abbott2017a, Abbott2017b, Abbott2017c} (which also led to the first estimate of the tidal deformability \cite{Abbott2018prlb, Abbott2019prx}), in addition to the precise NS mass measurements around $2 M_\odot$ --- PSR J1614-2230~\cite{Demorest2010}, PSR J0348$+$0432~\cite{Antoniadis2013}, and PSR J0740+6620~\cite{Cromartie2020} --- and the recent quantitative estimation of the mass--radius relation obtained from the Neutron-star Interior Composition ExploreR (NICER) measurements \cite{Riley2019,Riley2021, Raaijmakers2019, Miller2019, Miller2021, Salmi2022, Vinciguerra2024, Rutherford2024} have, in the last few years, pushed the study of dense-matter properties in NSs.
The increasing number of observations expected from the LIGO-Virgo-KAGRA collaboration (see Refs.~\cite{Ligo2015, Virgo2023, KAGRA:2021duu} and references therein), as well as those foreseen from the planned third-generation detectors and instruments like Einstein Telescope and Cosmic Explorer \cite{Maggiore2020, Branchesi2023, Evans2021}, are expected to provide valuable data from which new information on the structure and composition of NSs can be inferred.

To reliably connect astrophysical observations to the properties of dense matter in the NS interior, an equation of state (EoS), that is, the functional relation between the pressure and the (mass--)energy density, is needed. 
Indeed, general relativity guarantees a direct correspondence between static properties of cold beta-equilibrated NSs (which is a valid assumption for mature isolated NSs or coalescing NSs in their inspiral phase, as those considered in this manuscript) and the EoS.
Currently, the uncertainties in the EoS remain large, particularly in the high-density region. 
Indeed, for the outer crust (corresponding to the first few hundred metres under the NS surface) the only nuclear inputs in the calculation of the EoS and composition are nuclear masses (either experimentally measured or theoretically calculated) thus the model dependence is relatively small and only affects the deeper layers, where masses of neutron-rich nuclei have to be determined from theoretical mass models. 
In the inner crust (from the so-called neutron drip up to the crust–core transition at about $0.5 n_{\rm sat}$, the saturation density being $n_{\rm sat} \approx 0.16$~fm$^{-3}$) additional model dependencies in the EoS arise, because of the need to model the co-existence of neutron-rich clusters with a background of unbound neutrons (and electrons) (see e.g. Refs.~\cite{Haensel2007, Chamel2008, Oertel2017, Burgio2018, Blaschke2018} for a review, and references therein).
In the core, the possible appearance of non-nucleonic degrees of freedom significantly increases the spread in the EoS predictions (see e.g. Refs.~\cite{Haensel2007, Blaschke2018, Oertel2017, Raduta2021, Raduta2022}).
In view of these uncertainties, agnostic EoSs are often used in inference schemes.
These models do not rely on any specific description of the microphysics and are only subject to general physics constraints such as causality and thermodynamic stability.
Examples of these models are parametric EoS realisations (such as piecewise polytropes, spectral representations, or sound-speed models) and non-parametric ones (such as Gaussian processes).
Based on the idea that the low-density EoS is relatively well constrained and has little effect on the NS macroscopic quantities, and due to the difficulties in calculating the inhomogeneous crust, these EoSs are usually matched to a (unique) given crust (typically, the crust of Refs.~\cite{Baym1971, Douchin01}; see e.g., Refs.~\cite{Greif2019, Landry2019, Essick2020, Essick2021, Raithel2023, Huang2024}).
However, it has been shown that non-unified EoSs (meaning that different nuclear models are employed for the different regions of the NS) and the way the crust is matched to the core EoS can lead to spurious errors in the prediction of the NS macroscopic properties (see e.g. Refs.~\cite{Fortin2016, Ferreira2020, Suleiman2021, Davis2024}).

In Ref.~\cite{Davis2024}, we presented an efficient way to construct consistent and unified crust–core EoSs of cold and beta-equilibrated NSs for astrophysical (in particular gravitational-wave signal) analyses.
The corresponding numerical tool, the Crust (Unified) Tool for Equation-of-state Reconstruction (\texttt{CUTER}; hereby denoted \texttt{CUTER v1} for the first released version), is currently publicly accessible on the Zenodo repository\footnote{\url{https://zenodo.org/records/10781539}; \url{https://doi.org/10.5281/zenodo.10781539}.}.
This thus allows for an easy and open access by the astrophysical community and guarantees science reproducibility.

In this paper, we present the next release of the \texttt{CUTER} tool (hereby \texttt{v2}), currently available for the LIGO-Virgo-KAGRA collaboration and also available on the Zenodo repository\footnote{\url{https://zenodo.org/records/15166920}; \url{https://doi.org/10.5281/
zenodo.15166920}.}.
In this version, a new functionality has been added, aiming to reconstruct the NS outer crust.
Indeed, although the outer crust only accounts for the first few hundred metres in the NS, it is needed for the correct computation of the Tolman-Oppenhaimer-Volkoff (TOV) equations \cite{Tolman1939, Oppenheimer1939}.
Nevertheless, for some EoSs in the literature, the outer crust is not calculated at all, or it can exhibit jumps. 
These can naturally arise in a one-component plasma approach\footnote{In the one-component plasma approach, the NS outer crust is supposed to be divided in layers, each made of one nuclear species immersed in a gas of electrons.}, causing small discontinuities to appear at each change of composition.
In some cases, jumps can be due to thermodynamic inconsistencies in the calculation of the outer-crust EoS, thus inducing a non-physical non-monotonous behaviour in pressure or enthalpy (that is equal to the baryon chemical potential at zero temperature) or both.
These spurious effects can cause issues in the computation of the NS properties, in particular since TOV solvers often require monotonic pressure and enthalpy.
For these reasons, we added this new feature in the \texttt{CUTER} code.

The paper is organised as follows:
we present in Sect.~\ref{sec:reconstr} the numerical tool in its current version, describing both the whole-crust and outer-crust reconstruction in Sects.~\ref{sec:whole} and \ref{sec:outer}, respectively.
We then discuss the numerical results in Sect.~\ref{sec:results}, specifically the code validation and the applications of \texttt{CUTER v2} in Sects.~\ref{sec:code-valid} and \ref{sec:eos-app}, respectively. 
Finally, we present our conclusions in Sect.~\ref{sec:concl}.

\section{Equation-of-state reconstruction}
\label{sec:reconstr}

The aim of the proposed tool is to construct a unified and thermodynamically consistent EoS starting from a high-density beta-equilibrated EoS. 
Two functionalities are available: 
(i) the `whole-crust functionality', if the original EoS is provided only for the core, or for (part of) the inner crust and the core and it is not unified (that is, the crust and the core EoSs have been matched ad hoc); (ii) the `outer-crust functionality', if the original EoS for the inner crust and the core is unified, but the outer crust is missing.
These functionalities are described in Sects.~\ref{sec:whole} and \ref{sec:outer}, respectively.

From a practical point of view, the Python interface provided with the tool will automatically select the proper functionality to run, once the input EoS is specified\footnote{The users may also specify a fixed value of the baryon number density from which they wish to recompute the low-density EoS.}.
In addition, while the original \texttt{CUTER v1} only allowed for two input formats for the EoS, the \texttt{CompOSE} \cite{compose} and the \texttt{LAL} \cite{lalsuite} formats, it can now accommodate a `free format' EoS, meaning that the user can provide two of the following quantities (in nuclear units): the baryon number density, the energy density, and the pressure\footnote{If the user only provides the baryon number density and the energy density, or the energy density and the pressure, the other thermodynamic variable is recalculated through the thermodynamic relations; see App.~\ref{sec:app}.}.

\subsection{Whole-crust functionality}
\label{sec:whole}

The `whole-crust functionality' (outer plus inner-crust reconstruction) was already provided with \texttt{CUTER v1}.
It now operates if the original EoS is given starting at a baryon number density above $10^{-4}$~fm$^{-3}$.
The main idea is that, from an input beta-equilibrated EoS and a few isoscalar nuclear matter parameters, the tool allows one to obtain the isovector parameters and to consistently reconstruct the crust EoS, to be matched to the original EoS.
The EoS reconstruction is described in detail in Ref.~\cite{Davis2024}; we recall here the main points.

Starting from a relation between the energy density of cold beta-equilibrated NS matter $\mathcal{E}_\beta$ and the baryon number density $n_B$ in the zero-temperature approximation, it is possible to write the energy per baryon of homogeneous nucleonic (neutron, proton, and electron) matter that is consistent with the assumed $\mathcal{E}_\beta$ as\footnote{We use the uppercase $E$, the lowercase $e$, and $\mathcal{E}$ to denote the total energy, the energy per nucleon, and the energy per unit volume, respectively.}
\begin{eqnarray}
e_{\rm nuc,\beta}(n_B) &=& \frac{1}{n_B}\left [ \mathcal{E}_\beta(n_B) - \mathcal{E}_e(n_e(n_B)) \right. \nonumber \\
&& - \left. n_e m_p c^2 - (n_B - n_e) m_n c^2 \right] \ ,
\label{eq:enuc}
\end{eqnarray}
with $m_n$ and $m_p$ being the neutron and proton mass\footnote{The expression for the nucleon masses adopted for the subtraction in Eq.~\eqref{eq:enuc} (see also Eq.~\eqref{eq:edens-meta}) might not be exactly the same as that employed in the original EoS. This can in principle lead to a small inconsistency that cannot be avoided if details on the mass definition in the original EoS model are not known.}, respectively, $c$ the speed of light, and $n_e$ the electron density.
The latter quantity corresponds to the condition of neutrino-less beta equilibrium in the nucleonic regime among the chemical potentials:
\begin{equation}
\mu_n(n_B)-\mu_p(n_B)=\mu_e(n_e) \ ,
\label{eq:beta}
\end{equation}
where the electron chemical potential is simply given by its Fermi energy,
\begin{equation}
 \mu_e = \sqrt{(\hbar c)^2(3 \pi^2 n_e)^{2/3} + (m_e c^2)^2} \ ,
 \label{eq:mue}
\end{equation}
$\hbar$ and $m_e$ being the reduced Planck constant and the electron mass, respectively.
In Eqs.~\eqref{eq:enuc} and \eqref{eq:beta}, we made the assumption that only nucleonic degrees of freedom and no muons are present, which remains a safe hypothesis below saturation density.

The electron energy density, $\mathcal{E}_e$, is given by the expression for an ideal Fermi gas,
\begin{eqnarray}
\mathcal{E}_e &=& \frac{(m_e c^2)^4}{8 \pi^2 (\hbar c)^3}\left [ x_r(2x_r^2+1)\sqrt{x_r^2+1} \right. \nonumber \\
&-& \left. \ln \left( x_r + \sqrt{x_r^2+1} \right) \right ] \ ,
\label{eq:e_el}
\end{eqnarray}
with $x_r= \hbar c (3\pi^2 n_e)^{1/3}/(m_e c^2)$ being the relativity parameter (see e.g. Ref.~\cite{Weiss2004}). 

To consistently construct the crust EoS, the energy of nuclear matter, $e_{\rm nuc}(n_B,\delta)$, where ${\delta = (n_n-n_p)/n_B}$ is the isospin asymmetry parameter defined through the neutron and proton number densities, $n_n$ and $n_p$, needs to be known outside of the beta-equilibrium condition $\delta_\beta(n_B)$.
The latter quantity is related to the electron density through the neutrality conditions, that for a pure nucleonic content reads ${n_e=n_B\,(1-\delta_\beta)/2}$.
Following Ref.~\cite{Margueron2018}, we write the energy density of nucleonic matter in a meta-model approach:
\begin{equation}
 \mathcal{E}_{\rm nuc}(n_B,\delta) = 
 \sum_{q=n,p}n_q \, m_qc^2  + n_B e_{\rm nuc}(n_B,\delta)  \ ,
 \label{eq:edens-meta}
\end{equation}
where $q=n,p$ labels neutron and proton, respectively, and the energy per baryon is given by
\begin{equation}
e_{\rm nuc}(n_B,\delta) = t_{\rm FG}^\star(n_B,\delta) +e_{\rm is}(n_B)+e_{\rm iv}(n_B)\delta^2 \ ,
\label{eq:meta}
\end{equation}
where the kinetic term $t_{\rm FG}^\star$ includes the dominant deviation to the parabolic approximation\footnote{Indeed, because of this term, Eq.~\eqref{eq:meta} is not a purely parabolic approximation for the isospin dependence.} as well as the effective mass contribution, and the residual isoscalar $e_{\rm is}$ and isovector $e_{\rm iv}$ terms contain the most important model dependence (see Eq.~\eqref{eq:e0}).

In order to equate Eqs.~\eqref{eq:enuc} and \eqref{eq:meta} along the beta-equilibrium line, thus finding $\delta=\delta_\beta(n_B)$, the following equation has to be solved (see Sect.~2.1 in Ref.~\cite{Davis2024} for details):
\begin{eqnarray}
&& 2 \left. \frac{\partial t_{\rm FG}^\star}{\partial \delta}\right|_{n_B}(n_B,\delta_\beta) - \mu_e(n_B,\delta_\beta) + \Delta m_{np} c^2 \nonumber \\
&& = -\frac{4}{n_B \delta_\beta}\left [ \mathcal{E}_\beta(n_B) - \mathcal{E}_e(n_e) \right. \nonumber \\ 
&& - n_e m_p c^2 - (n_B - n_e) m_n c^2 \nonumber \\
&&  \left. - n_B t_{\rm FG}^\star(n_B,\delta_\beta) - n_B e_{\rm is}(n_B) \right] \ ,
\label{eq:final}
\end{eqnarray}
where $\Delta m_{np} = m_{n} - m_{p}$, and we have used Eq.~\eqref{eq:beta} and the definition of the neutron and proton chemical potentials,
\begin{eqnarray}
\label{eq:muq}
 \mu_q &=& \left. \frac{\partial \mathcal{E}_{\rm nuc}}{\partial n_q} \right|_{n_{q^\prime}} 
 = e_{\rm nuc}+m_q c^2 + \left. n_B \frac{\partial e_{\rm nuc}}{\partial n_B} \right|_\delta \nonumber \\
 &+& (-\delta \pm 1) \left. \frac{\partial e_{\rm nuc}}{\partial \delta}\right|_{n_B} \ , 
 \end{eqnarray}
where the plus (minus) sign holds for neutrons (protons).
For each value of $n_B$, Eq.~\eqref{eq:final} can be numerically solved for $\delta_\beta$, as proposed in Ref.~\cite{Essick2021} (see their Eq.~(22)), although the expression is slightly more involved because of the explicit inclusion of the kinetic term in Eq.~(\ref{eq:meta}).

The solution of Eq.~\eqref{eq:final} requires knowledge of the isoscalar energy functional $e_{\rm is}(n_B)$.
For the latter, we used the meta-model labelled `ELFc' in Ref.~\cite{Margueron2018}, truncated at order $N$,
\begin{equation}
e_{\rm is}(n_B)=\sum_{k=0}^N \frac{v_k^{\rm is}}{k!}x^k u_k(x) \ ,
\label{eq:e0}
\end{equation}
where $x=(n_B-n_{\rm sat})/(3n_{\rm sat})$, the function $u_k(x)$,
\begin{equation}
u_k(x)=1-(-3x)^{N+1-k} e^{-b n_B/n_{\rm sat}} \ ,
\label{eq:uk}
\end{equation}
with $b=10\ln 2$ \cite{Margueron2018}, ensures the correct zero-density limit, and the parameters $v_k^{\rm is}$ are directly connected to the so-called isoscalar nuclear empirical parameters (see Eqs.~(18)-(22) in Ref.~\cite{Davis2024}). 
Indeed, it was shown that, when truncating the expansion at order $N=4$ (and even $N=3$), the meta-model gives a very good reproduction of realistic functionals at low density (see Refs.~\cite{Margueron2018, Davis2024} for a discussion on the truncation order).
Moreover, a simple harmonic approximation as proposed in Ref.~\cite{Essick2021} is correct around saturation, but it is not sufficient (see e.g. Fig.~5 in Ref.~\cite{Margueron2018}) if one wants to extract the functional behaviour at very low density.
This is of particular importance since the low-density region is the domain better constrained by the ab initio calculations, and it was shown to be also influential for a correct calculation of the crust \cite{Dinh2021a}. 

Once the isoscalar parameters at order $N$ ($n_{\rm sat}$, $E_{\rm sat}$, $K_{\rm sat}$, $Q_{\rm sat}$, $Z_{\rm sat}$ for $N=4$) are known, Eq.~(\ref{eq:final}) is solved for $N+1$ different sub-saturation density points, $n_{B,j}$, corresponding to $N+1$ points $x_j$, $j=1\dots N+1$ (see Table 1 in Ref.~\cite{Davis2024}). 
The isovector empirical parameters ($E_{\rm sym}, L_{\rm sym}, K_{\rm sym},Q_{\rm sym}$, $Z_{\rm sym}$ for $N=4$) can thus be analytically obtained by matrix inversion \cite{Mondal2022, Davis2024}.
This `inversion procedure' thus allows one to extract the nucleonic functional $e_{\rm nuc}(n_B, \delta)$ (see Eqs.~\eqref{eq:enuc} and \eqref{eq:meta}) from a given beta-equilibrated EoS $\mathcal{E}_\beta(n_B)$ and a set of isoscalar parameters\footnote{Whenever the isoscalar empirical parameters are not provided or known, it is possible to assign specific values to them. In both \texttt{CUTER v1} and \texttt{v2} we propose that these parameters are fixed to those of the BSk24 functional ($n_{\rm sat}=0.1578$~fm$^{-3}$, $E_{\rm sat}=-16.048$~MeV, and $K_{\rm sat}=245.5$~MeV; see Ref.~\cite{Goriely2013}) in order to perform the inversion procedure at the lowest order $N=2$.}.

All isoscalar and isovector parameters being known, the crust EoS and the crust--core transition point are calculated as originally described in Refs.~\cite{Carreau2019, Dinh2021b, Dinh2021c} (see also Ref.~\cite{Davis2024}) within a compressible liquid-drop model (CLDM) in a one-component-plasma approach.
Specifically, at each layer of the crust, characterised by a pressure $P$ and baryon density $n_B$, matter is modelled as a periodic lattice consisting of Wigner-Seitz cells of volume $V_{\rm WS}$ containing one type of cluster with $Z$ protons of mass $m_p$ and $A-Z$ neutrons of mass $m_n$ ($A$ being the cluster total mass number), immersed in a uniform gas of electrons and, in the inner crust, of neutrons as well. 
For a given thermodynamic condition, defined by the baryonic number density $n_B$, the total energy density of the inhomogeneous system can thus be written as
\begin{equation}
    \mathcal{E}_{\rm inhom} = \mathcal{E}_e +  \mathcal{E}_{\rm g} (1-u) + \frac{E_i}{V_{\rm WS}} \ ,
    \label{eq:energy-density}
\end{equation}
where $\mathcal{E}_e(n_e)$ is the electron gas energy density, $\mathcal{E}_{\rm g} = \mathcal{E}_{\rm nuc}(n_{\rm g}, \delta_{\rm g}) $ is the neutron-gas energy density (including the rest masses of nucleons) at baryonic density $n_{\rm g}$ and isospin asymmetry $\delta_{\rm g} = 1$ (no protons are present in the gas at $T=0$), and $u=\frac{A/n_i}{V_{\rm WS}}$ is the volume fraction of the cluster with internal density $n_i$. Finally, the third term on the right hand side of Eq.~(\ref{eq:energy-density}), $- \mathcal{E}_{\rm g} u$, accounts for the interaction between the cluster and the neutron gas, that we treated in the excluded-volume approach.
In the outer crust, $\mathcal{E}_{\rm g}$ is set to zero.
The last term in Eq.~(\ref{eq:energy-density}) is the cluster energy per Wigner-Seitz cell: 
\begin{equation}
    E_i = M_i c^2 + E_{\rm bulk} + E_{\rm Coul} + E_{\rm surf +  curv} \ ,
    \label{eq:Ei}
\end{equation}
where $M_i = (A-Z)m_n + Zm_p$ is the total bare mass of the cluster, $E_{\rm bulk} = A e_{\rm nuc}(n_i, 1-2Z/A)$ is the cluster bulk energy, and $E_{\rm Coul} + E_{\rm surf + curv} = V_{\rm WS} (\mathcal{E}_{\rm Coul} + \mathcal{E}_{\rm surf} + \mathcal{E}_{\rm curv})$ accounts for the total interface energy, that is, the Coulomb interaction between the nucleus and the electron gas (including the proton--proton, proton--electron, and electron--electron interactions) as well as the residual interface interaction between the nucleus and the surrounding dilute nuclear-matter medium.
For the bulk energy density of the ions, $\mathcal{E}_{\rm bulk} = e_{\rm nuc}n_i$, and that of the neutron gas, we use the same functional expression, that is, the meta-model approach of Ref.~\cite{Margueron2018} (see Eq.~\eqref{eq:edens-meta}).
We consider here only spherical clusters, since the so-called pasta phases that may appear at the bottom of the inner crust are expected to have only a small impact on the EoS (see Ref.~\cite{Dinh2021a} for a discussion on pasta phases).
The Coulomb energy density thus reads\footnote{The last term of Eq.~\eqref{eq:Fcoul} corresponds to the lattice energy that can be equivalently expressed as
\begin{equation*}
\mathcal{E}_L = -\frac{9}{10} \frac{e^2 Z^2}{r_N} \frac{u^{1/3}}{V_{\rm WS}} \ .    
\end{equation*} 
In this work and in Ref.~\cite{Davis2024}, we used the Madelung constant $\approx 0.896$ \cite{Haensel2007} instead of the coefficient $0.9$.}
\begin{eqnarray}
    \mathcal{E}_{\rm Coul}  &=& \frac{2}{5}\pi (e n_i r_N)^2 u \left(\frac{1-I}{2}\right)^2 \nonumber \\
    && \left[ u+ 2  \left( 1- \frac{3}{2}u^{1/3} \right) \right] \ , 
    \label{eq:Fcoul}
\end{eqnarray}
with $e$ being the elementary charge, $I = 1-2Z/A$, and $r_N = (3A/(4 \pi n_i))^{1/3}$.
For the surface and curvature contributions, we employed the same expression as in Refs.~\cite{Maruyama2005,Newton2013}, that is
\begin{equation}
\mathcal{E}_{\rm {surf}} + \mathcal{E}_{\rm {curv}} =\frac{3u}{r_N} \left( \sigma_{\rm s}(I) +\frac{2\sigma_{\rm c}(I)}{r_N} \right) \ , 
\label{eq:interface}   
\end{equation}
where $\sigma_{\rm s}(I)$ and $\sigma_{\rm c}(I)$ are the surface and curvature tensions \cite{Ravenhall1983},
\begin{eqnarray}
\sigma_{\rm s}(I) &=& \sigma_0 \frac{2^{p+1} + b_{\rm s}}{y_p^{-p} + b_{\rm s} + (1-y_p)^{-p}} \ , \label{eq:sigma0_1} \\
\sigma_{\rm c} (I) &=& 5.5 \sigma_{\rm s}(I) \frac{\sigma_{0, {\rm c}}}{\sigma_0} (\beta -y_p) \ ,
\label{eq:sigma0_2}
\end{eqnarray}
where $y_p = (1-I)/2$ and the surface parameters $(\sigma_0, \sigma_{0, {\rm c}}, b_{\rm s}, \beta)$ were optimised for each set of bulk parameters and effective mass to reproduce the experimental nuclear masses in the 2020 Atomic Mass Evaluation (AME) table \cite{ame2020}, while we set $p=3$ \cite{Carreau2019b}. 
The EoS and composition of the crust are then obtained by variationally minimising the energy density of the Wigner-Seitz cell with $(A, I, n_i, n_e, n_{\rm g})$ as variational variables, under the constraint of baryon number conservation and charge neutrality holding in every cell \cite{Carreau2019, Dinh2021a, Dinh2021b, Davis2024}.

Once the low-density part of the EoS is computed, the original EoS is matched either at the crust--core transition consistently calculated with the \texttt{CUTER} code described above\footnote{The crust--core transition is determined by comparing the energy per baryon of the inhomogeneous crust to that of the homogeneous neutron-proton-electron matter in the core.}, or at the first point of the original (high-density) EoS table, if the latter is after the crust--core transition. 
We note that, since the nuclear empirical parameters used for the crust are consistent with those used for the core EoS around saturation density, the (crust and core) EoS is, in this sense, unified, and a continuous and smooth behaviour in the energy density is ensured.

\subsection{Outer-crust functionality}
\label{sec:outer}

The `outer-crust functionality' is a new feature of \texttt{CUTER v2}.
It allows to compute the outer crust if the original EoS is missing it, or if the one provided in the input EoS exhibits discontinuities or thermodynamic inconsistencies.
Indeed, as discussed in the introduction, the outer crust is needed for the correct computation of the NS global properties such as NS radii (see Fig.~\ref{fig:m-r-d1mstar} and the related discussion in Sect.~\ref{sec:code-valid}).
This functionality runs when the EoS provided by the user starts at a baryon number density below $10^{-4}$~fm$^{-3}$.
The outer crust is then reconstructed using analytical representations of the Brussels-Montreal Skyrme (BSk) EoSs based on the BSk energy-density functionals (either BSk22 or BSk24, the latter being recommended) \cite{Goriely2013, Pearson2018, Pearson2018err}\footnote{The analytical expressions of the fits to the tabulated EoSs are coded in the \texttt{CUTER v2}. They are also available in Ref.~\cite{Pearson2018} and the Fortran routines allowing for the computation of the EoS and composition are available on the Ioffe website \url{http://www.ioffe.ru/astro/NSG/BSk/index.html}.}, that naturally smooth out possible jumps and garantees a monotonic behaviour in pressure and enthalpy. 

The original EoS is then matched either at the point in the baryon number density which ensures thermodynamic consistency between the outer-crust and the original EoS (following the Gibbs condition, this point is defined as the point where $P_{\rm oc} = P_{\rm ic}$ and $\mu_{B, {\rm oc}} = \mu_{B, {\rm ic}}$, `oc' and `ic' denoting the outer and inner crust, respectively) or at the lowest entry point of the original EoS table if a thermodynamically consistent matching is not possible.
An option for replacing the entire outer crust with the chosen BSk one up to the limiting density of the outer crust in that model is available, too.  
We fixed by default the lower limiting density to $10^{-11}$~fm$^{-3}$, to comply with a surface density as low as possible for numerical reasons in some TOV solvers.
However, it has to be noted that below about $10^{-10}$~fm$^{-3}$, temperature effects might influence the EoS (see Ref.~\cite{Haensel2007}), whereas we continue to assume zero temperature for computing the EoS.

We present the performance of the code for the two functionalities in Sect.~\ref{sec:results}.

\subsection{Neutron-star properties}
\label{sec:NS-prop}

Once the (beta-equilibrated) EoSs $P(\mathcal{E})$ are calculated, the properties of a non-rotating NS are computed from the TOV equations \citep{Tolman1939, Oppenheimer1939}: 
 \begin{eqnarray}
     \label{eq:TOV1}
     \frac{{\rm d}P(r)}{{\rm d}r} &=& -\frac{G \mathcal{E}(r)\mathcal{M}(r)}{c^2 r^2}
     \biggl[1+\frac{P(r)}{\mathcal{E}(r)}\biggr] \nonumber \\ 
     && \times \biggl[1+\frac{4\pi P(r)r^3}{c^2\mathcal{M}(r)}\biggr]\biggl[1-\frac{2G\mathcal{M}(r)}{c^2 r}\biggr]^{-1} \ ,
     \end{eqnarray}
where $G$ is the gravitational constant and
 \begin{equation}
     \label{eq:TOV2}
     \mathcal{M}(r) = \frac{4\pi}{c^2} \int_0^r 
     \mathcal{E}(r')r'^2{\rm d}r' \ .
\end{equation}
The integration of Eqs.~\eqref{eq:TOV1}-\eqref{eq:TOV2} until the surface of the star defines its radius $R$ and mass $M=\mathcal{M}(r=R)$\footnote{The surface of the star is typically defined in terms of a lower limit on the density. For this reason, some TOV solvers require the EoS to start from a very low density, see also Sect.~\ref{sec:outer} for a brief discussion.}.
In addition, the dimensionless tidal deformability $\Lambda = \lambda(r=R)$ is calculated as
\begin{equation}
   \lambda(r) = \frac{2}{3} k_2(r) \left[\frac{rc^2}{G\mathcal{M}(r)}\right]^5 \ , 
 \label{eq:lambda} 
\end{equation} 
with the tidal Love number function $k_2(r)$ given by~\citep{Hinderer2008, Hinderer2010}
\begin{align}
     k_2 =& \,\frac{8C^5}{5}(1-2C)^2\big[2+2C(y-1)-y\big]\nonumber \\
    &\times \Big\{2C\big[6-3y+3C(5y-8)\big] \nonumber \\ 
    &+ 4C^3\big[13-11y+C(3y-2)+2C^2(1+y)\big] \nonumber \\ 
     &+ 3(1-2C)^2\big[2-y+2C(y-1)\big] \ln (1-2C) \Big\}^{-1} \ ,
     \label{eq:k2}
\end{align}
where the dimensionless compactness parameter reads
\begin{equation}
    C(r) = \dfrac{G\, \mathcal{M}(r)}{r\, c^2}\, ,
\end{equation}
and the function $y(r)$ is obtained by integrating the following differential equation with the boundary condition $y(0)=2$:
\begin{equation}
     \label{y_eq}
     r \dfrac{dy}{dr}+ y(r)^2 + F(r) y(r) + Q(r)=0 \, ,
\end{equation}
\begin{equation}
     \label{F(r)}
     F(r)=\frac{1-4\pi G r^2(\mathcal{E}(r)-P(r))/ c^4}{1-2G\mathcal{M}(r)/(r c^2)}\, ,
\end{equation}
\begin{align}
     \label{Q(r)}
     Q(r)=&\frac{4 \pi G r^2/c^4}{1-2G\mathcal{M}(r)/(r c^2)}\Biggl[5\mathcal{E}(r)+9P(r)\nonumber \\
     &+\frac{\mathcal{E}(r)+P(r)}{c_s(r)^2} c^2-\frac{6\,  c^4}{4\pi r^2 G}\Biggr] \nonumber \\
     &-4\Biggl[ \frac{G(\mathcal{M}(r)/(r c^2)+4\pi r^2 P(r)/c^4)}{1-2G\mathcal{M}(r)/(r c^2)}\Biggr]^2\, ,
\end{align}
with $ c_s=c \sqrt{dP/d\mathcal{E}}$ denoting the sound speed.

\section{Numerical results}
\label{sec:results}

We present here the numerical results obtained within the framework described in Sect.~\ref{sec:reconstr}.
We first present the \texttt{CUTER v2} code validation in Sect.~\ref{sec:code-valid} and its application to some known EoSs in Sect.~\ref{sec:eos-app}.

\subsection{\texttt{CUTER v2} code validation}
\label{sec:code-valid}

In order to validate the procedure and illustrate the results, we considered different EoSs available on the \texttt{CompOSE} database \cite{Typel2015, Typel2022, compose}.
We start by discussing the code validation for the following EoSs: RG(SLY4)\footnote{\url{https://compose.obspm.fr/eos/134}} \cite{Gulminelli2015} and GPPVA(DDME2)\footnote{\url{https://compose.obspm.fr/eos/218}} \cite{Grill2012, Grill2014}. 
The former EoS is based on the non-relativistic Skyrme-type SLy4 \cite{Chabanat1998} energy-density functional, while the latter one is based on the relativistic energy-density functional DDME2~\cite{Lalazissis2005} with density-dependent couplings.
The choice of these models was driven by the fact that their nuclear-matter parameters are consistent with current knowledge of nuclear physics and they both yield EoSs that are able to support $2 M_\odot$ NSs. 
Moreover, the EoSs based on those functionals are also in agreement with constraints from GW170817 (see e.g. Fig.~3 in Ref.~\cite{Dinh2021a}).

\begin{figure*}[ht]
\centering
\includegraphics[scale=0.4]{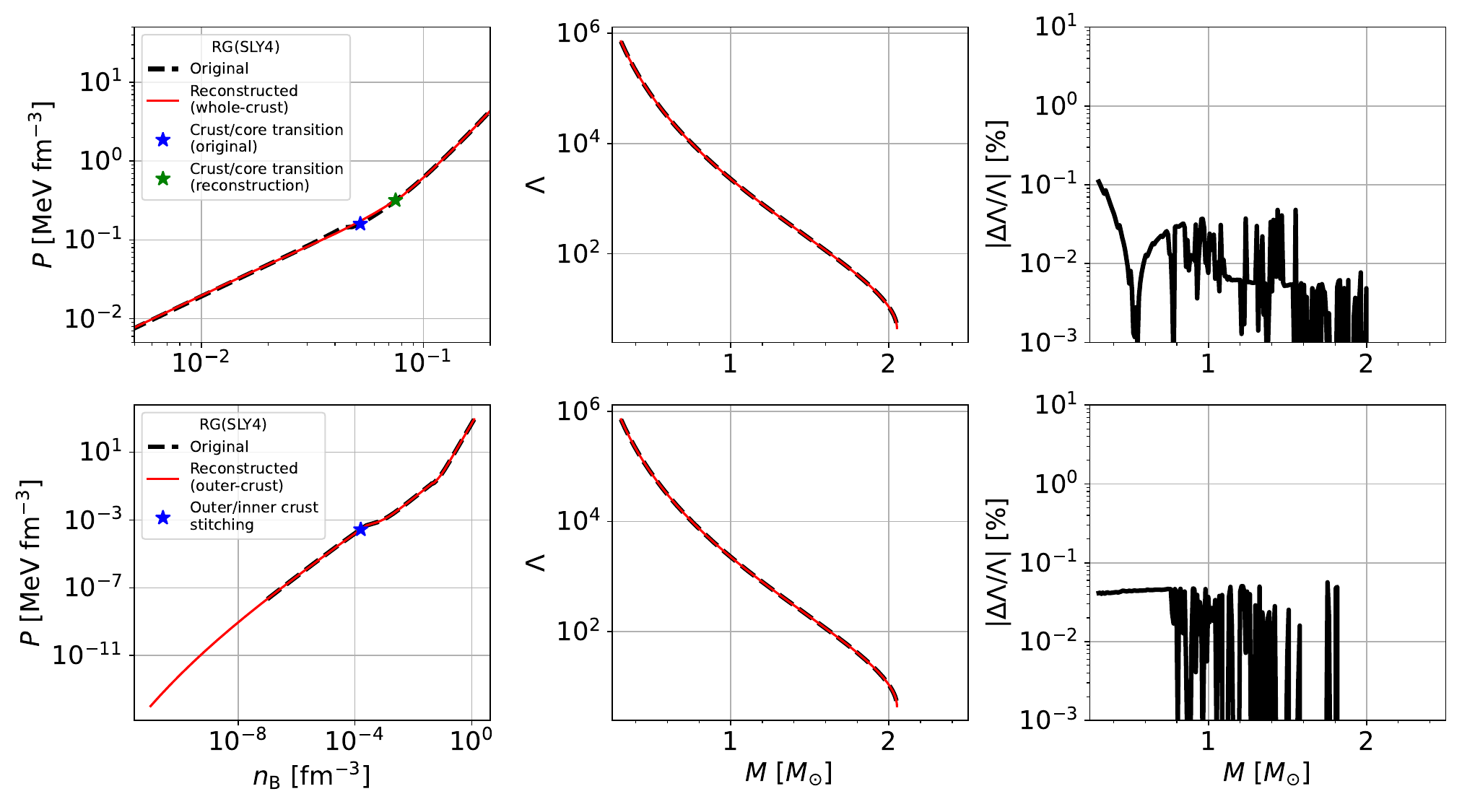}
\caption{Pressure versus baryon number density (left panels), $\Lambda$-mass relation (middle panels), and absolute relative error on the $\Lambda$ (right
panels) for the whole-crust reconstruction (top panels) and the outer-crust reconstruction (bottom panels), for the RG(SLY4) EoS. Black dashed lines (left and middle panels) correspond to the original EoS, while red solid lines correspond to the reconstructed EoS. Stars indicate the crust-core boundary (top panel) and the outer-inner crust boundary (bottom panel); see text for details. }
\label{fig:eos-rg}
\end{figure*}

\begin{figure*}[ht]
\centering
\includegraphics[scale=0.4]{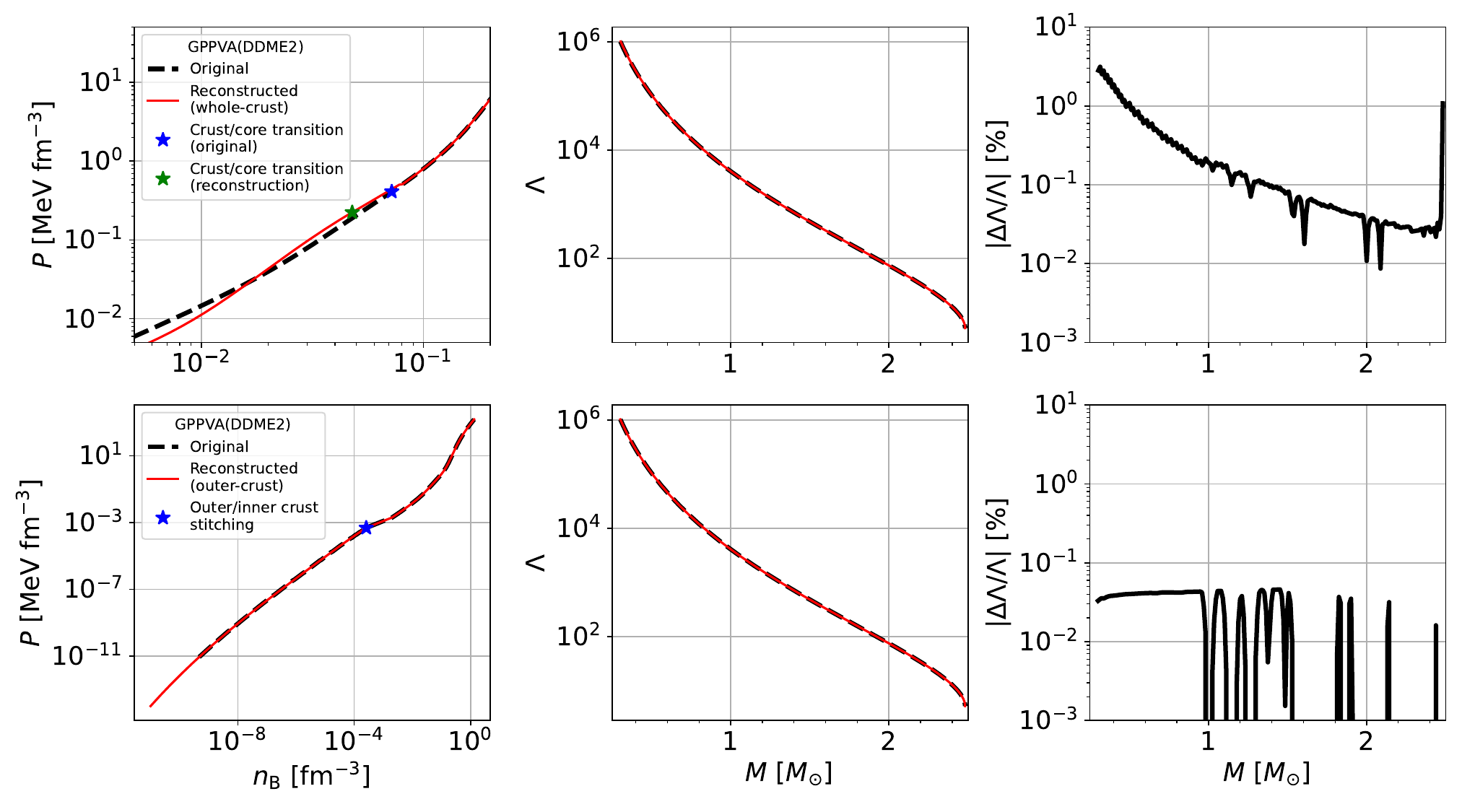}
\caption{Same as in Fig.~\ref{fig:eos-rg} but for the GPPVA(DDME2) EoS.}
\label{fig:eos-gp}
\end{figure*}

We start the discussion by showing the results obtained using the whole-crust functionality.
This feature has been discussed in detail in Ref.~\cite{Davis2024}, but for completeness we show here some complementary results.
This demonstrates that the new added feature to the code still keeps the expected performance of the original functionality, too.
The results of the reconstruction, that was performed at order 3 (see Ref.~\cite{Davis2024} for a more thorough discussion about the order of the reconstruction), are shown in the top panels of Figs.~\ref{fig:eos-rg} and \ref{fig:eos-gp}.
The pressure versus baryon number density in the inner crust--core transition region is shown in the left panels: dashed black lines correspond to the original EoS table, while red solid lines correspond to the reconstructed EoS.
Green (blue) stars mark the crust--core transition of the reconstructed (original) EoS.
As already discussed in Ref.~\cite{Davis2024}, the agreement with the original EoS can be considered satisfactory (see also Table 3 in Ref.~\cite{Davis2024}).
Indeed, the values of the crust--core transition density and pressure depend not only on the underlying functional, but also on the many-body method used to calculate them.
For example, the crust--core transition predicted by the RG(SLY4) \citep{Gulminelli2015} and the DH(SLy4) \citep{Douchin01} EoSs, both based on the SLy4 functional, are $n_{\rm cc}=0.052$~fm$^{-3}$ and $P_{\rm cc}=0.16$~MeV~fm$^{-3}$ (shown in the top left panel of Fig.~\ref{fig:eos-rg} by the blue star), and $n_{\rm cc}=0.076$~fm$^{-3}$ and $P_{\rm cc}=0.34$~MeV~fm$^{-3}$, respectively, while we obtain $n_{\rm cc}=0.075$~fm$^{-3}$ and $P_{\rm cc}=0.32$~MeV~fm$^{-3}$ for the reconstructed EoS (green star in the top left panel of Fig.~\ref{fig:eos-rg}).
As for the GPPVA(DDME2) EoS, we obtain for the reconstructed EoS $n_{\rm cc} = 0.048$~fm$^{-3}$ and $P_{\rm cc}=0.22$~MeV~fm$^{-3}$ (green star in the top left panel of Fig.~\ref{fig:eos-gp}).
It has to be noted that the calculations of Ref.~\cite{Grill2014} (for which the crust--core transition is diplayed by the blue star in the top left panel of Fig.~\ref{fig:eos-gp}) include pasta phases, while in the present calculations only spheres are considered.
In Ref.~\cite{Dinh2021a}, it was shown that including pasta phases in the same CLDM approach as that employed in this work considerably improves the agreement with the value of $n_{\rm cc} = 0.07$~fm$^{-3}$ obtained in Ref.~\cite{Grill2014} (see Table~3 in Ref.~\cite{Dinh2021a}).
Neglecting non-spherical phases in the EoS reconstruction may thus (at least partially) account for the discrepancy observed with respect to the original model.

The corresponding reconstructed nuclear matter parameters are given for completeness in Table~\ref{tab:param}: the first lines correspond to the parameter set of the original model\footnote{In the \texttt{CompOSE} database the isoscalar parameters are given up to order 3 and the isovector ones are given up to order 2. The value of $Q_{\rm sym}$ shown in Table~\ref{tab:param} was taken from Refs.~\cite{Margueron2018, Dinh2021a}.}, while the following lines indicate the (input) isoscalar parameters and the reconstructed isovector ones at the order of the reconstruction ($N=3$ for both RG(SLY4) and GPPVA(DDME2)). 
As already discussed in Ref.~\cite{Davis2024}, the inversion procedure yields a good reproduction of the low-order parameters, while larger deviations are observed for the high-order ones.
This is because the points for the reconstructions are chosen around saturation where the higher order parameters are less influential thus the differences with respect to the original model arise from the capability of the polynomial expansion to reproduce the functional behaviour away from saturation.

To evaluate the impact on more global NS properties, we display in the top middle panels of Figs.~\ref{fig:eos-rg} and \ref{fig:eos-gp} the tidal deformability versus mass relation. 
We can see that the $\Lambda-M$ curves are almost indistinguishable, and that the relative error, displayed in the top right panels, remain below about $0.1\%$ for NS masses above $1\ M_\odot$ for both cases considered here, and reach $\sim 1\%$ for the GPPVA(DDME2) EoS for very low-mass stars.
This, on the one hand, confirms the results obtained in Ref.~\cite{Davis2024}, and, on the other hand, assures that the performance of \texttt{CUTER v1} is preserved.

We now turn to the discussion of the reconstruction of the EoS using the outer-crust functionality described in Sect.~\ref{sec:outer}.
The reconstructed EoSs, obtained employing the analytical representation of the EoS based on the BSk24 functional for the outer crust, are compared to the original EoSs for RG(SLY4) and GPPVA(DDME2) in the bottom panels of Figs.~\ref{fig:eos-rg} and \ref{fig:eos-gp}, respectively. 
Stars indicate the matching point between the constructed outer-crust and the original EoS. 
For RG(SLY4), we use the thermodynamically consistent matching option while for GPPVA(DDME2) we apply the option that replaces the entire outer crust. 
As one can see, the agreement with the original EoS is very satisfactory (left panels).
To evaluate the impact on the NS properties, we plot the tidal deformability versus mass (middle panels) and the corresponding relative errors (right panels).
We can observe that the tidal deformability is very well reproduced, also for low-mass stars, where the crust is expected to contribute the most, showing that the outer-crust-reconstruction procedure does not introduce spurious effects or downgrade the reproduction of the NS global properties.
We note that the outer crust based on the BSk22 functional was implemented in the GPPVA(DDME2) EoS on the \texttt{CompOSE} database.
We can see from the bottom panels of Fig.~\ref{fig:eos-gp} that the difference in the EoS and in the computed NS tidal deformability (and, in general, global NS observables) between the EoS using the `original' (BSk22 based) and the reconstructed (BSk24 based) outer crust remains small as long as a thermodynamically consistent `reasonable' outer crust is implemented.
Indeed, the difference on the $\Lambda$ is below about $0.05\%$ for the considered EoSs for the shown NS mass range. 
We note that large parts of the outer crust are well determined by nuclear measurements, such that as soon as the EoS is consistently constructed, there is not much model dependence. 
In addition, as mentioned earlier, the outer crust only has a sub-dominant effect on global NS properties, such that the exact choice of the nuclear model for the outer crust reconstruction does not impact very significantly the results. 
Indeed, for RG(SLY4), the baryon number density and pressure at the matching point between the reconstructed outer crust and the original EoS are respectively $1.4\times{10}^{-4}$~fm$^{-3}$ and $2.3\times{10}^{-4}$~MeV~fm$^{-3}$ for the BSk22-based outer-crust EoS, and $1.6\times{10}^{-4}$~fm$^{-3}$ and $2.8\times{10}^{-4}$~MeV~fm$^{-3}$ for the BSk24-based outer-crust EoS. 
Concerning the relative error in the tidal deformability, values are less than $0.1\%$ for all of the considered NS masses, irrespective of the BSk model employed for the outer-crust reconstruction.

\begin{figure*}[ht]
\centering
\includegraphics[scale=0.4]{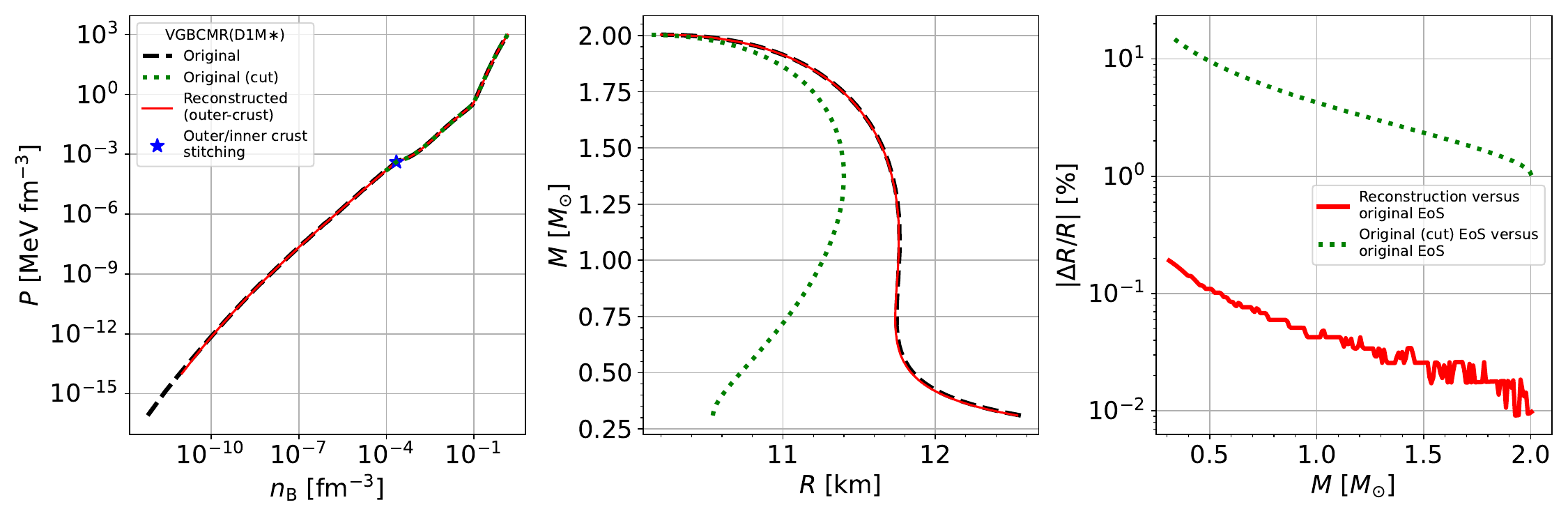}
\caption{Pressure versus baryon number density (left panel), mass-radius relation (middle panel), and absolute relative error on the radius (right panel) for the outer-crust functionality for the VGBCMR(D1M$^{\star}$) EoS. Black dashed lines (left and middle panels) correspond to the original EoS and the green dotted curves show the original `cut' EoS. 
For comparison, the reconstructed EoS is displayed with red solid curves.
In the left panel, the blue star indicates where the outer crust has been stitched to the cut EoS; see text for details.}
\label{fig:m-r-d1mstar}
\end{figure*}

On the other hand, the complete neglect of the outer crust has an important impact, particularly for the computation of the radii.
This is illustrated in Fig.~\ref{fig:m-r-d1mstar} for the VGBCMR(D1M$^\star$)\footnote{\url{https://compose.obspm.fr/eos/254}} \cite{Mondal2020, Vinas2021} EoS, based on the non-relativistic Gogny D1M$^\star$ effective interaction \cite{Gonzalez2018}. 
The pressure versus baryon number density for this EoS is shown in the left panel. 
Two cases are considered: the original EoS (black dashed line), and the same EoS but where the outer crust has been removed (that is, the EoS was artificially `cut' at $n_{\mathrm{B}}\approx{10^{-4}}$~fm$^{-3}$; green dotted line). 
The corresponding mass--radius curves and the absolute relative error between the original and cut cases are shown in the middle and right panels, respectively.
For a $1.4 M_{\odot}$ NS, the lack of an outer crust can give errors in the calculated radius by a few $\%$, and up to $\approx 10\%$ for very low mass NSs.
Although relatively small, these errors are of the same order of the precision expected from current and future astrophysical observations.
Moreover, completely neglecting the outer crust notably changes the $M-R$ curve.
For comparison, we also show in Fig.~\ref{fig:m-r-d1mstar} the quality of the outer-crust reconstruction performed by \texttt{CUTER v2} (red solid lines), obtained employing the analytical representation of the BSk24 EoS for the outer crust, that was matched at the lowest entry point of the cut table.
One can see that the radius is well reproduced. 
Indeed, the absolute relative error in radius between the original and reconstructed EoS (right panel) is less than about $0.1\%$ for $M > 0.5 M_\odot$ NSs.

\subsection{Application of \texttt{CUTER} to existing non-unified EoSs. Example cases.}
\label{sec:eos-app}

\begin{figure*}[ht]
\centering
\includegraphics[scale=0.3]{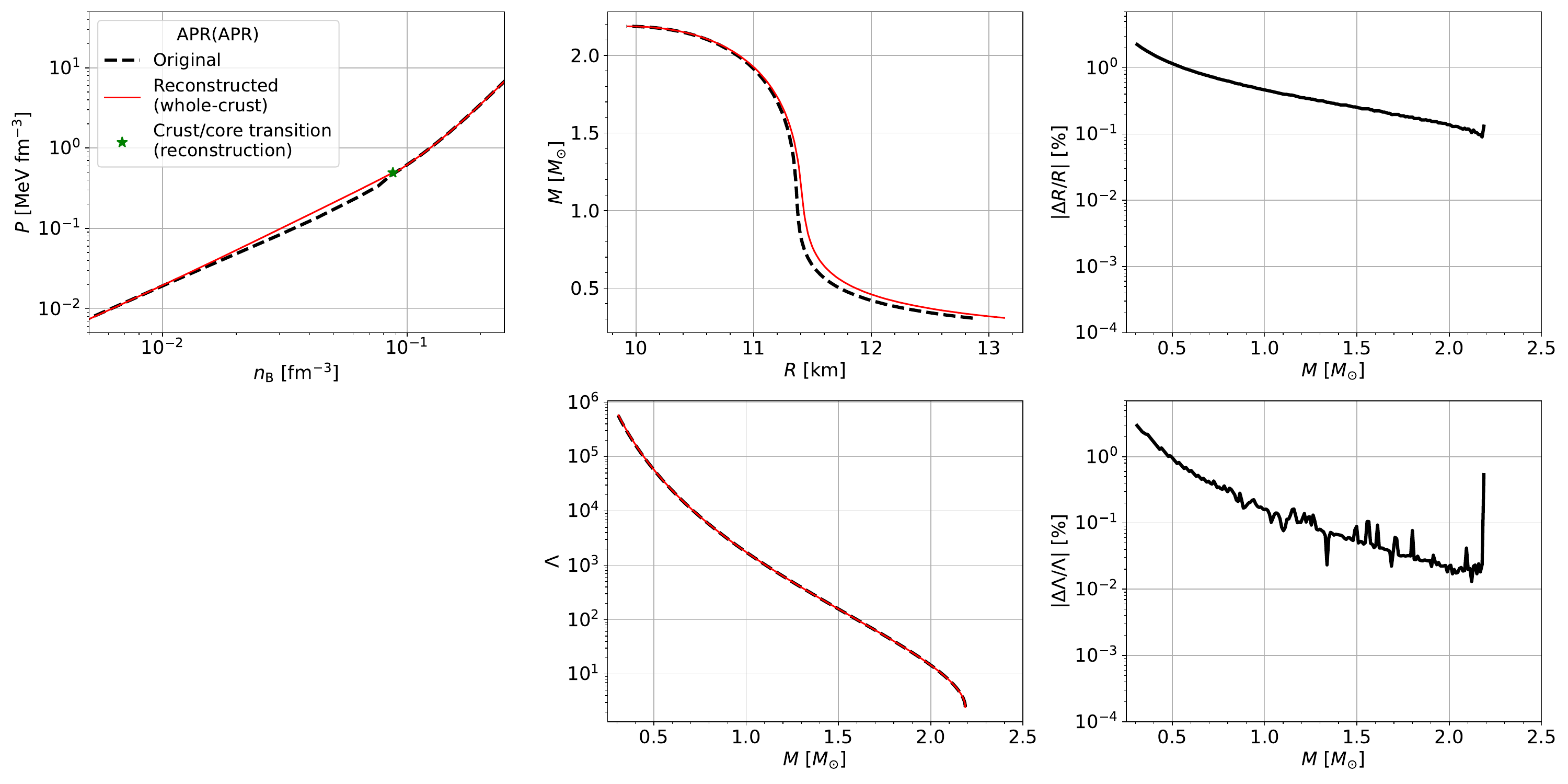}
\caption{Top panels: pressure versus baryon number density (left panel), mass-radius relation (middle panel), and absolute relative differences on the radius (right panel) for the whole-crust reconstruction for the APR(APR) EoS from \texttt{CompOSE}. Bottom panels: $\Lambda$-mass relation (middle) and absolute relative difference on $\Lambda$ (right panel). Symbols and curve legends are the same as in the top panels of Fig.~\ref{fig:eos-rg}; see text for details.}
\label{fig:m-r-apr}
\end{figure*}

We discuss now the application of \texttt{CUTER} to some existing non-unified EoSs.
As a first example we consider the APR EoS \cite{APR1998} which is widely employed in the context of NS physics. 
It contains a transition to a phase with a pion condensate appearing at densities $\gtrsim 0.2$ fm$^{-3}$. 
In the most commonly used version of that EoS, this transition is modelled via an inhomogeneous mixed phase with a Gibbs construction, see the detailed explanation in Chap.~5.12 of Ref.~\cite{Haensel2007}. 
This EoS, as non-unified version, is available as APR(APR) EoS from \texttt{CompOSE}\footnote{\url{https://compose.obspm.fr/eos/68}}. 
The outer crust is thereby taken from Ref.~\cite{Baym1971,Haensel1994} (Baym-Pethick-Sunderland, BPS) and the inner crust uses the SLy4 model of Ref.~\cite{Douchin01}. For this EoS, we performed the whole-crust reconstruction at order 2 using the nuclear empirical parameters shown in Table~\ref{tab:param} under the label `APR(APR)', which are also the ones given in the \texttt{CompOSE} database. 
The corresponding EoS, $M-R$ and $\Lambda - M$ curves for the original and reconstructed EoSs are displayed in Fig.~\ref{fig:m-r-apr}. 
The baryon number density and the pressure at the crust--core transition from the reconstruction procedure are $n_{\rm cc} = 0.087$~fm$^{-3}$ and $P_{\rm cc} = 0.49$~MeV~fm$^{-3}$, respectively (marked as a green star in the top left panel of Fig.~\ref{fig:m-r-apr}).
These values are not so far from the baryon number density and pressure where the homogeneous APR(APR) EoS was attached to the non-unified crust, that is, 0.08~fm$^{-3}$ and 0.38~MeV~fm$^{-3}$, respectively.
The relative differences in the NS radius and $\Lambda$ with respect to the non-unified EoS are less than about $2\%$ for the range of considered masses, as can be seen from the right panels of Fig.~\ref{fig:m-r-apr}.
Although relatively small, these differences are avoidable by employing a unified model, as the one provided by \texttt{CUTER}.

\begin{figure*}[ht]
\centering
\includegraphics[scale=0.3]{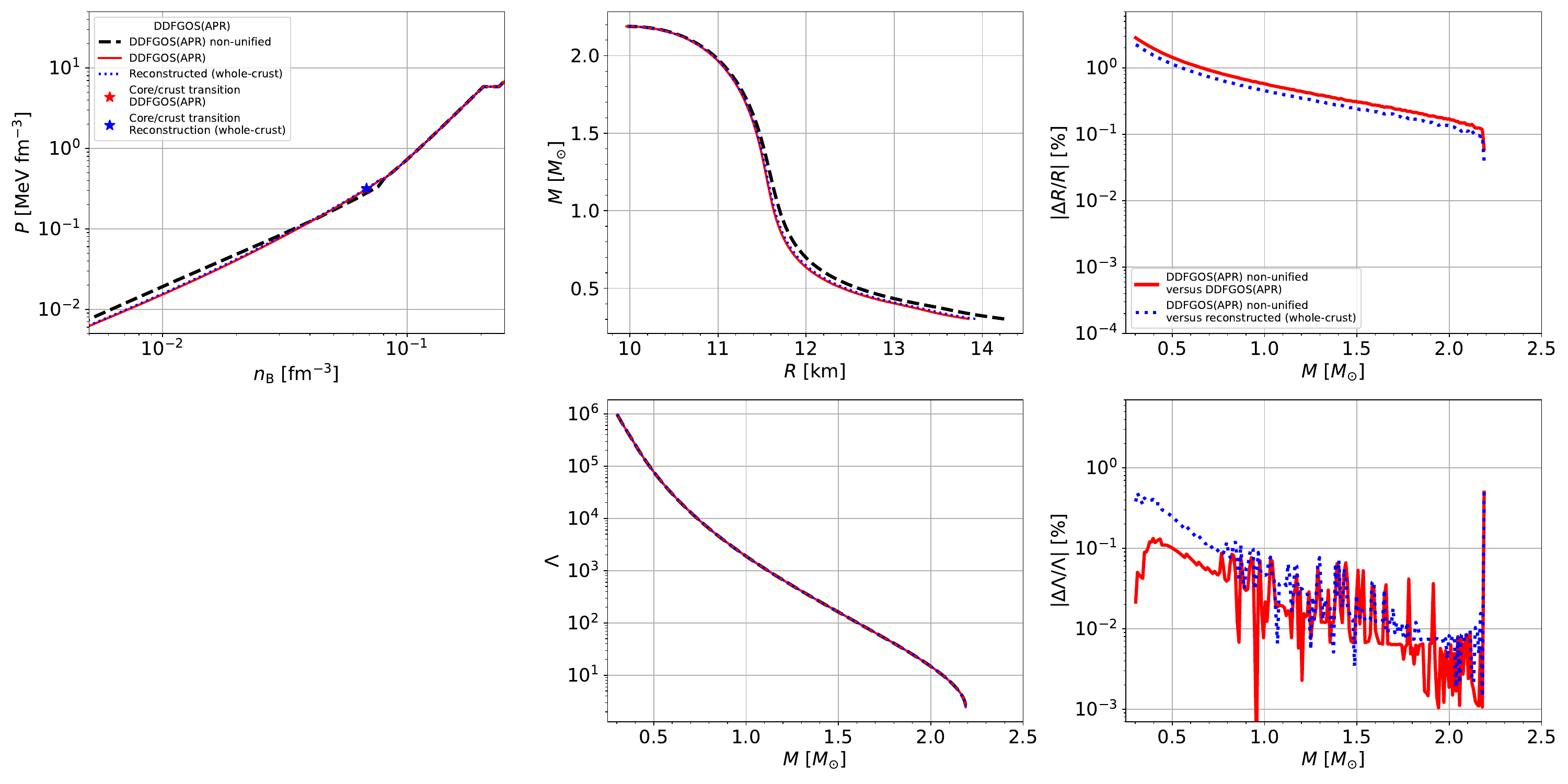}
\caption{Same as in Fig.~\ref{fig:m-r-apr} but for DDFGOS(APR) EoS. In the left and middle panels, the red solid and blue dotted curves correspond to the whole-crust reconstructions, while the black dashed curve corresponds to the non-unified EoS. In the right panels, the errors are given with respect to the non-unified EoS; see text for details.}
\label{fig:eos-ddfgos}
\end{figure*}

As an alternative, we propose a version of the APR EoS, that we denote as `DDFGOS(APR)', where the transition to the pion condensed phase is handled as a first-order phase transition with two homogeneous phases following a Maxwell construction of constant pressure and baryon number chemical potential across the transition:
\begin{equation}
P_{\mathrm{NP}} = P_{\mathrm{PC}} \ ; \quad \mu_{B, \mathrm{NP}} = \mu_{B,\mathrm{PC}} \ ,
\label{eq:Maxwell}
\end{equation}
where `NP' corresponds thereby to the purely nucleonic phase at low densities and `PC' to the phase with a pion condensate. 
An ideal gas of charged electrons and muons is present in both phases with $\mu_e = \mu_\mu = \mu_n - \mu_p$ determined via electrical charge neutrality and beta equilibrium. 
We note that the lepton number chemical potential $\mu_L$ and the charge chemical potential $\mu_Q$ do not enter the phase equilibrium conditions, Eq.~\eqref{eq:Maxwell}, since, first, neutrino-less beta equilibrium in both phases imposes $\mu_{L,\mathrm{NP}} = \mu_{L,\mathrm{PC}} \equiv 0$. 
Second, due to the electrical charge neutrality condition, assuming that both phases are composed of homogeneous matter, the Coulomb interaction exactly vanishes and the associated chemical potential $\mu_Q$ becomes ill-defined \cite{Ducoin2007,Gulminelli2013}.   
For this EoS, the homogeneous nucleonic phase is described within the framework of Ref.~\cite{Constantinou:2014hha} and presents some differences in the corresponding density region with respect to the APR(APR) EoS. The nuclear matter parameters are slightly different, too, from those of the APR(APR) and are listed in Table~\ref{tab:param}. 
The unified reconstructed EoS is thus calculated with this given set of parameters (both the isoscalar and isovector ones, at order 2) and shown in the top left panel of Fig.~\ref{fig:eos-ddfgos} (red solid line). 
We find the crust-core transition point at the following values of baryon number density and pressure: $n_{\rm cc} = 0.068$~fm$^{-3}$ and $P_{\rm cc} = 0.31$~MeV~fm$^{-3}$ (marked as a red star in the figure), which are slightly lower than those of the APR(APR) EoS.
To check the performance of \texttt{CUTER} for this EoS, we have also tested the reconstruction of the isovector parameters, that are listed in Table~\ref{tab:param} as `Order 2', and calculated the corresponding reconstructed EoS, which is shown in Fig.~\ref{fig:eos-ddfgos} as a blue dotted line\footnote{To this aim, we artificially set the isovector parameters listed in Table~\ref{tab:param} to zero to force the parameter reconstruction by \texttt{CUTER}.}.
We can see that the nuclear empirical parameters and the EoS are reproduced with high accuracy; indeed, the two curves (red solid and blue dotted lines) and the crust-core transition points (stars) are almost indistinguishable in the figure.
Finally, to compare the DDFGOS(APR) EoS with a non-unified one, we follow the same procedure as for the `original' APR(APR) one on \texttt{CompOSE}, that is, we attached the same non-unified low-density crust of Refs.~\cite{Baym1971,Haensel1994,Douchin01} below $0.08$~fm$^{-3}$.
This latter EoS is displayed as a black dashed line in the top left panel of Fig.~\ref{fig:eos-ddfgos}.
The corresponding $M$--$R$ and $\Lambda$--$M$ relations for these EoSs are displayed in the middle panels of Fig.~\ref{fig:eos-ddfgos}.
We can see that the curves corresponding to the two unified DDFGOS(APR) EoSs are almost identical, and that the differences observed between the unified and the non-unified DDFGOS(APR) are reflected in the global NS quantities.
The absolute difference in the tidal deformability amounts to less than $0.1\%$ for NSs with $M \gtrsim 0.5 M_\odot$, while the discrepancy in the radius is globally larger and amounts to less than $1\%$ in the same NS-mass range (see right panels of Fig.~\ref{fig:eos-ddfgos}). 

\begin{figure*}[ht]
\centering
\includegraphics[scale=0.3]{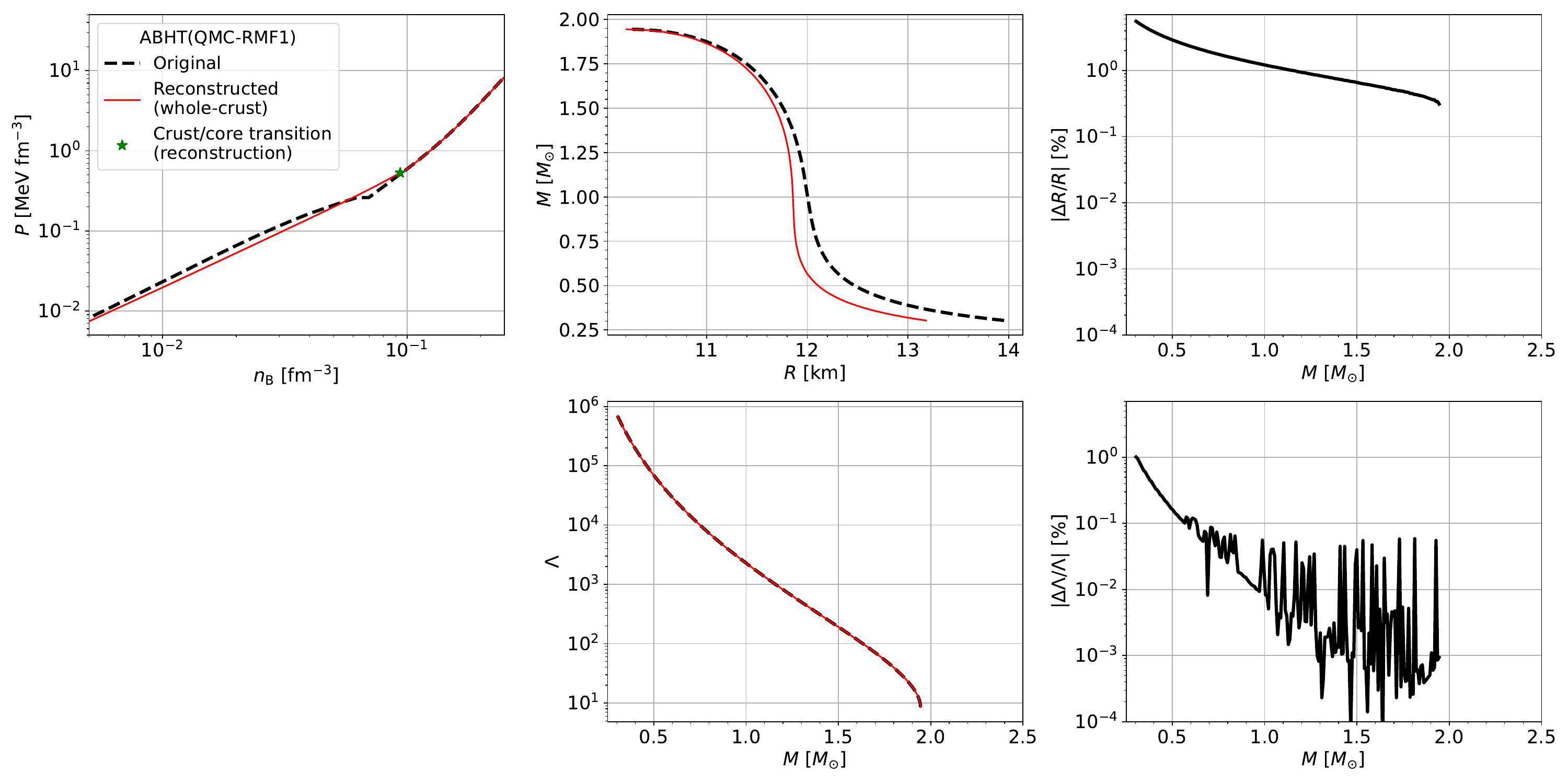}
\caption{Same as Fig.~\ref{fig:m-r-apr} but for the ABHT(QMC-RMF1) EoS.}
\label{fig:m-r-qmc}
\end{figure*}

As an additional example, we consider the EoS ABHT(QMC-RMF1)\footnote{\url{https://compose.obspm.fr/eos/275}} \cite{Alford2022}, also available on the \texttt{CompOSE} database. 
This EoS consists of an homogeneous nucleonic core starting at 0.07~fm$^{-3}$ and is calculated using a relativistic mean-field model constrained by chiral effective field theory calculations of pure neutron matter as well as properties of isospin-symmetric nuclear matter around saturation density. 
As for the APR EoS, the model is non unified, that is, the outer and inner crusts have been matched.
For ABHT(QMC-RMF1) EoS, the outer and inner crust were constructed respectively using the Baym-Pethick-Sunderland \cite{Baym1971} and the GPPVA(TM1e) \cite{Grill2014} EoSs, and the core starts at a baryon number density of 0.07~fm$^{-3}$ and a pressure of 0.263 MeV fm$^{-3}$. 
The results of the whole-crust reconstruction (at order 3) are shown in Fig.~\ref{fig:m-r-qmc} and the corresponding reconstructed nuclear parameters are given in Table~\ref{tab:param}. 
We can see that the nuclear empirical parameters are recovered with good accuracy and the reconstructed EoS is smoother.
The crust--core transition baryon number density and pressure that we obtain are $n_{\rm cc} = 0.094$~fm$^{-3}$ and $P_{\rm cc} =0.53$~MeV~fm$^{-3}$, respectively.
These values are higher than the point where the crust has been originally matched to the core, yielding a larger relative difference in the $M-R$ curve with respect to the APR(APR) case. 
The relative differences between the original and reconstructed values of the NS radii are indeed up to a few $\%$ for low-mass NSs, and remain below $1\%$ for the $\Lambda$ (see right panels in Fig.~\ref{fig:m-r-qmc}).
We point out again that the \texttt{CUTER} reconstructed EoS should yield more reliable results for the NS observables since they are consistently computed with a unified EoS.

The three newly constructed unified versions of the APR(APR), the DDFGOS(APR), and ABHT(QMC-RMF1) EoS are available on \texttt{CompOSE}\footnote{The unified version of APR(APR) can be found at \url{https://compose.obspm.fr/eos/328}, the DDFGOS(APR) at \url{https://compose.obspm.fr/eos/327}, and the unified version of ABHT(QMC-RMF1) at \url{https://compose.obspm.fr/eos/326}.}.

\begin{table*}[]
    \centering
    \caption{Nuclear empirical parameters for the RG(SLY4), GPPVA(DDME2), APR(APR), DDFGOS(APR), and ABHT(QMC-RFT1) EoSs. The first and second rows in each case give the original and reconstructed parameters, respectively. The saturation density $n_{\mathrm{sat}}$ is given in units of fm$^{-3}$ while the other nuclear empirical parameters are given in units of MeV.}
    \begin{tabular}{c | c c c c c c c c}
        \hline
        \hline
            & $n_{\mathrm{sat}}$ & $E_{\mathrm{sat}}$ & $K_{\mathrm{sat}}$ & $Q_{\mathrm{sat}}$ 
            & $E_{\mathrm{sym}}$ & $L_{\mathrm{sym}}$ & $K_{\mathrm{sym}}$ & $Q_{\mathrm{sym}}$ \\ 
            \hline
            RG(SLY4) set & 0.159 & -15.97 & 230.0 & -363.11 & 32.0 & 46.0 & -119.7 & 521.0 \\ 
            Order 3 & 0.159 & -15.97 & 230.0 & -363.11 & 31.8 & 44.3 & -139.5 & 310.6 \\ 
        \hline
        GPPVA(DDME2) set & 0.152 & -16.14 & 251.0 & 479.0 & 32.3 & 51.0 & -82.7 & 777 \\ 
        Order 3 & 0.152 & -16.14 & 251.0 & 479.0 & 32.5 & 51.0 & -107.9 & 496.2 \\ 
        \hline
            APR(APR) set & 0.16  & -16.0 & 266.0 & - & 32.6 & 57.6 & - & - \\ 
            Order 2 & 0.16 & -16.0  & 266.0 & - & 32.7 &     37.5  & -222.1 & - \\ 
        \hline
            DDFGOS(APR) set & 0.16  & -16.0 & 266.0 & - & 32.59 & 58.47 & -102.63 & - \\ 
            Order 2 & 0.16  & -16.0 & 266.0 & - & 32.91 & 58.89 & -104.29 & - \\ 
        \hline
        ABHT(QMC-RMF1) set & 0.16 & -16.1 & 260.0 & -496.0 & 32.9 & 44.5 & -191.0 & - \\ 
        Order 3 & 0.16 & -16.1 & 260.0 & -496.0 & 32.6 & 39.9 & -183.9 & 826.0 \\ 
        \hline
    \end{tabular}
    \label{tab:param}
\end{table*}

\section{Conclusions}
\label{sec:concl}

In this work, we have presented the \texttt{CUTER v2} tool, aiming at providing unified and thermodynamically consistent EoSs for astrophysical applications, like gravitational-wave analyses.
The current version provides two functionalities: (i) the `whole-crust functionality', introduced in Ref.~\cite{Davis2024}, that reconstructs a unified and consistent EoS starting from a high-density beta-equilibrated EoS, and (ii) the `outer-crust functionality', that adds a consistent outer-crust EoS to a unified inner-crust and core EoS for which the outer-crust is missing or thermodynamically inconsistent.
The code has been mainly tested on nucleonic EoSs, but can be in principle applied to any EoS.
We show that the code is able to efficiently perform both tasks, allowing the computation of NS global properties in a consistent way.




\bmhead{Acknowledgements}
The authors would like to thank N. Stergioulas for fruitful comments on the code and the manuscript.
This material is based upon work supported by NSF's LIGO Laboratory which is a major facility fully funded by the National Science Foundation. The authors gratefully acknowledge the Italian Instituto Nazionale di Fisica Nucleare (INFN), the French Centre National de la Recherche Scientifique (CNRS), and the Netherlands Organization for Scientific Research for the construction and operation of the Virgo detector and the creation and support of the EGO consortium.

\section*{Declarations}
\begin{itemize}
\item Funding. This work has been partially supported by the IN2P3 Master Project NewMAC and MAC, the ANR project `Gravitational waves from hot neutron stars and properties of ultra-dense matter' (GW-HNS, ANR-22-CE31-0001-01), the CNRS International Research Project (IRP) `Origine des \'el\'ements lourds dans l'univers: Astres Compacts et Nucl\'eosynth\`ese (ACNu)', the National Science Foundation (grant Number PHY 21-16686). 
\item Conflict of interest. Not applicable. 
\item Ethics approval and consent to participate. Not applicable.
\item Data availability. All data generated or analysed during this study are included in this published article. The EoSs presented in Sect.~\ref{sec:eos-app} are available on the \texttt{CompOSE} website: the unified version of APR(APR) can be found at \url{https://compose.obspm.fr/eos/328}, the DDFGOS(APR) at \url{https://compose.obspm.fr/eos/327}, and the unified version of ABHT(QMC-RMF1) can be found at \url{https://compose.obspm.fr/eos/326}.
\item Materials availability. Not applicable.
\item Code availability. The \texttt{CUTER v2} code is available at \url{https://zenodo.org/records/15166920}, \url{https://doi.org/10.5281/
zenodo.15166920}.
\item Author contribution. All authors contributed equally to this work.
\end{itemize}

\noindent

\begin{appendices}

\section{Expressions used for the recalculation of thermodynamic variables}
\label{sec:app}

In \texttt{CUTER v2}, the user can provide the input EoS in three formats: \texttt{CompOSE}, \texttt{LAL}, and a `free format'. 
The latter format allows for two cases: the user can provide either the baryon number density $n_B$ and the energy density of beta-equilibrated matter $\mathcal{E}_\beta$, or the pressure $P$ and the energy density $\mathcal{E}_\beta$\footnote{For the free format, the electron chemical potential and the electron fraction cannot be obtained for the input EoS by \texttt{CUTER} and will be put to zero in the output format.}.
In the first case, the baryon chemical potential $\mu_B$ and the pressure $P$ are reconstructed.
To reconstruct the pressure, the first law of thermodynamics is used,
\begin{equation}
    P(\mathcal{E}_{\beta}, n_B) = n_B^2 \frac{\rm d}{{\rm d} n_B} \left( \frac{\mathcal{E}_{\beta}}{n_B} \right) \ . 
    \label{eq:P-rec}
\end{equation}

In the second case, if only $\mathcal{E}_\beta$ and $P$ are provided, first the log-enthalpy $h$ is calculated from thermodynamic consistency as
\begin{equation}
    h(\mathcal{E}_{\beta}, P) = \int_0^P \frac{{\rm d}P^\prime}{\mathcal{E}_{\beta} + P^\prime} \ . \label{eq:h-rec} 
\end{equation}
The user can fix the integration constant by providing a value of the mass $m_B$ at the command line prompt. 
We recommend that a value of $m_B = m_u$ be used (with $m_u$ the atomic mass unit). 

The baryon number density $n_B$ is then recomputed via the log-enthalpy $h$ as
\begin{equation}
    n_B(\mathcal{E}_{\beta}, P) = \frac{1}{m_B}  e^{-h} \left( \mathcal{E}_{\beta} + P \right) \ . 
    \label{eq:nb-rec}
\end{equation}

In both cases, the baryon chemical potential is computed from the pressure, energy density and baryon number density as
\begin{equation}
    \mu_{B,\beta} = \frac{\mathcal{E}_\beta + P}{n_B} \ .
    \label{eq:mu-rec}
\end{equation}

\end{appendices}


\bibliography{biblio}

\end{document}